\documentclass[twocolumn,english,aps,prl,showpacs,superscriptaddress,groupedaddress]{revtex4}
\usepackage[T1]{fontenc}
\usepackage[latin9]{inputenc}
\usepackage{amsmath}
\usepackage{graphicx}
\usepackage{amssymb}

\makeatletter
\@ifundefined{textcolor}{}
{%
 \definecolor{BLACK}{gray}{0}
 \definecolor{WHITE}{gray}{1}
 \definecolor{RED}{rgb}{1,0,0}
 \definecolor{GREEN}{rgb}{0,1,0}
 \definecolor{BLUE}{rgb}{0,0,1}
 \definecolor{CYAN}{cmyk}{1,0,0,0}
 \definecolor{MAGENTA}{cmyk}{0,1,0,0}
 \definecolor{YELLOW}{cmyk}{0,0,1,0}
 }

\usepackage{dcolumn}
\usepackage{bm}
\usepackage{dcolumn}

\makeatother

\usepackage{babel}

\begin{document}

\title{Sub-Rayleigh lithography using high flux loss-resistant entangled
states of light}

\author{Shamir Rosen, Itai Afek, Yonatan Israel, Oron Ambar, Yaron Silberberg}

\affiliation{Department of Physics of Complex Systems, Weizmann Institute of Science,
Rehovot, Israel}
\begin{abstract}
Quantum lithography achieves phase super-resolution using fragile,
experimentally challenging entangled states of light. We propose a
scalable scheme for creating features narrower than classically achievable,
with reduced use of quantum resources and consequently enhanced resistance
to loss. The scheme is an implementation of interferometric lithography
using a mixture of an SPDC entangled state with intense classical
coherent light. We measure coincidences of up to four photons mimicking
multiphoton absorption. The results show a narrowing of the interference
fringes of up to $30\%$ with respect to the best analogous classical
scheme using only $10\%$ of the non-classical light required for
creating NOON states.
\end{abstract}


\maketitle
\emph{Introduction}. Quantum entanglement has been shown to be instrumental
in surpassing the classical limits in various technological fields
\cite{Gerry}. In the case of optical lithography, it has been proposed
that multi-photon entanglement can be harnessed to generate arbitrary
patterns with a higher resolution than that attainable with classical
light \cite{boto,Bjork,Kok}. Specifically, in the technique of interferometric
lithography proposed by Boto et. al \cite{boto}, a prescribed pattern
is written periodically on a multi-photon absorbing substrate via
interference between two intersecting beams. The entangled states
most often associated with quantum lithography are the so called 'high-NOON'
states, which are path-entangled Fock states with N-photons in either
one of two paths \cite{Boyd,dowling08}. Such states collect phase
N times faster than classical light, hence their super-resolution
capabilities. Indeed, efforts to generate high-NOON states have been
reported over the last decade \cite{D'Angelo,Steinberg,Zeilinger,afek},
striving for larger N values.

In interferometric lithography the minimal feature size is determined
by the minimal width that can be written with an interference pattern,
hence the drive to form narrower fringes. A Mach-Zehnder interferometer
is prototypical for studying such interferences. While a classical
source with a standard intensity detector will show a bright fringe
half a wavelength wide, it is possible to use N-photon absorption
(or, alternatively, an N-photon detector) to get a $\sqrt{N}$ narrower
feature, still using a classical source. In contrast, using such detectors
with a perfect N-photon NOON state would generate fringes that are
$N$ times narrower. The additional $\sqrt{N}$ narrowing due to the
entanglement of the light is a purely quantum effect.

\begin{figure}
\centering \includegraphics[width=7cm]{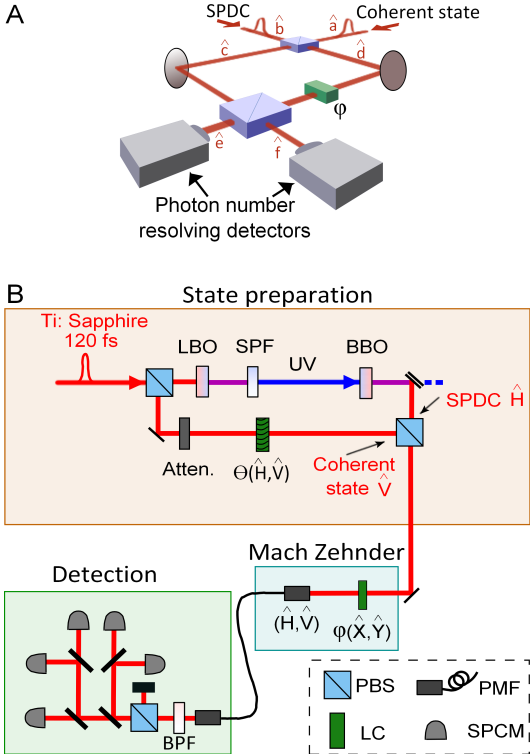} \caption{(color online). Experimental Setup. (A) Illustration of the generation
of the loss-resistant states in a Mach Zehnder (MZ) interferometer
fed by a coherent state and SPDC. The loss-resistant states are created
in modes $\hat{c}$ and $\hat{d}$ of the MZ and are subsequently
interfered and measured in a photon number resolving apparatus. (B)
Detailed layout of the setup. 120-fs pulses from a Ti:sapphire oscillator
operated at 80 MHz are up-converted using a 2.74-mm lithium triborate
(LBO) crystal, short pass filtered, and then down-converted using
a 1.78-mm beta barium borate (BBO) crystal, creating correlated photon
pairs at the original wavelength (808 nm). This SPDC ($\hat{H}$ polarization)
is mixed with attenuated coherent light ($\hat{V}$ polarization)
on a polarizing beamsplitter (PBS). A thermally induced drift in the
relative phase is corrected every few minutes with the use of a liquid
crystal (LC) phase retarder. The MZ is polarization-based in a collinear
inherently phase-stable design. The MZ phase is controlled using an
additional LC phase retarder at $45^{0}$, which adjusts the phase
between $\hat{X}$ and $\hat{Y}$ polarizations. The spatial and spectral
modes are matched using a polarization-maintaining fiber (PMF) and
a 3-nm (full width at half max) bandpass filter (BPF). Photon number-resolving
detection is performed using an array of single-photon counting modules
(SPCM, Perkin Elmer).}

\label{setup}
\end{figure}

Although NOON states possess exceptional super-resolution capabilities,
they are limited by heavy experimental constraints. All schemes for
generating high-NOON states use nonclassical light sources such as
spontaneous parametric down-converted (SPDC) light as a resource for
entanglement. For example, Fig.\ref{setup}(A) shows schematically
our approach, where an SPDC beam is mixed on a beamsplitter with a
coherent state to form an almost perfect superposition of NOON states
\cite{afek,Hofmann}. It is the brightness of the nonclassical source
that poses the main limitation in such systems, and it is important
to find ways to minimize the use of this resource.

Another experimental difficulty that comes hand-in-hand with the quantum
nature of the light is an exceptionally high sensitivity to loss.
NOON states in particular are N times more sensitive to loss compared
to classical pulses. This optical loss is practically unavoidable
as it is inherent to most optical elements as well as to any detection
system.

One possible path towards overcoming these difficulties is the use
of partially entangled states. Such states have recently been used
to achieve optimal sensitivity in phase measurements \cite{Walmsley-1}.
Here we introduce and generate a set of partially entangled states
designed for use in super-resolution lithography. We show theoretically
and demonstrate experimentally, that when an SPDC beam is mixed in
with a coherent state, significant narrowing of the N-photon fringe
occurs with significantly less SPDC than required for forming a NOON
state. Therefore, quantum resolution enhancement can be obtained with
reduced demand on the quantum resources. Moreover, we show that the
reduced use of SPDC enables a major increase in the generating flux
leading to high multi-photon absorption rates, even in the presence
of high losses. Another benefit we discovered is that these high flux
loss-resistant entangled states exhibit just one major fringe per
wavelength, whereas the other fringes that would appear in a NOON
state reduce in magnitude and hence will be easier to eliminate.

We note that another approach towards raising the quantum flux used
an unseeded OPA as the source of entangled photons. This scheme however
is limited to doubling the density of the interference fringes \cite{Agarwal,Sciarrino}.

\emph{Theoretical analysis}. We present here a detailed analysis of
a general two-mode entangled state created by mixing SPDC with a coherent
state and compute its interference pattern in a Mach Zehnder interferometer.
We find the dependence of the width of the interference fringes on
the parameters of the two input modes, and observe that minute amounts
of SPDC account for most of the fringe narrowing. Consider the setup
of Fig.\ref{setup}(A) where a Mach-Zehnder (MZ) interferometer is
fed by a coherent state, $|\alpha\rangle_{a}$, and SPDC,$|\xi\rangle_{b}$,
in its input ports. The lower path amplitude accumulates a controllable
phase shift $\varphi$ before the two modes interfere. Rather than
interfering on an absorbing substrate as originally suggested, modes
$\hat{c}$ and $\hat{d}$ impinge on two ports of a second beamsplitter
and are subsequently detected by photon number resolving detectors
$\textbf{e}$ and $\textbf{f}$. The event $|0,N\rangle_{e,f}$ is
equivalent to an N-photon absorption event \cite{boto}.

The input state can be written as

\begin{align}
|\psi\rangle_{a,b}=|\alpha\rangle_{a}\otimes|\xi\rangle_{b}\label{eq:set-eqs}\end{align}
 where $\alpha$ is a coherent state defined by:\\
 \begin{align}
|\alpha\rangle=\sum_{n=0}^{\infty}exp(-\frac{1}{2}|\alpha|^{2})\frac{\alpha^{n}}{\sqrt{n!}}|n\rangle,\end{align}
 \begin{align}
\alpha=|\alpha|exp(i\theta_{cs})\end{align}
 and $\xi$ is single mode degenerate SPDC with the following wave
function\cite{Gerry}: \begin{align}
|\xi\rangle=\frac{1}{\sqrt{coshr}}\sum_{m=0}^{\infty}(-1)^{m}\frac{\sqrt{(2m)!}}{2^{m}m!}(tanhr)^{m}|2m\rangle\end{align}

The controllable parameters are the amplitudes of the two input modes,
and the relative phase between them. We denote the pair amplitude
ratio of the coherent state and SPDC inputs $\gamma\equiv|\alpha|^{2}/r$
(in the limit $r,|\alpha|\ll1$)\cite{afek}. The pair amplitude ratio
$\gamma$ reflects the relative intensity of the two input states;
for a given coherent state, a smaller value of $\gamma$ reflects
stronger SPDC component hence more quantum resource.

We consider N-photon absorption, which, for a given state, has a probability
of $|\langle0,N|\psi\rangle_{e,f}|^{2}$. When the loss can be neglected,
the system conserves the total number of photons in the two-mode state.
This means that only input states with a total number of N photons
contribute to the N photon signal. We treat the subspace of N=3 as
an instructive example. The input state, before the first BS is:

\begin{equation}
|\psi_{3}\rangle_{a,b}=\frac{\alpha^{2}}{\sqrt{6}}|3,0\rangle_{a,b}-\frac{\alpha r}{\sqrt{2}}|1,2\rangle_{a,b}\end{equation}

Under the transformation of the first BS this becomes:

\begin{align}
|\psi_{3}\rangle_{c,d} & =\frac{\alpha}{4}(\frac{\alpha^{2}-3r}{\sqrt{3}}(|3,0\rangle_{c,d}+|0,3\rangle_{c,d})+\\
 & (\alpha^{2}+r)(|2,1\rangle_{c,d}+|1,2\rangle_{c,d}))\nonumber \end{align}

Note that the choice of $\alpha^{2}=-r$ would generate a perfect
NOON state; we continue to consider the general case. When passing
through a phase shifter (PS), number states collect a phase that is
equal to the PS phase, $\varphi,$multiplied by the number of photons
in the state. Explicitly:

\begin{align}
|\psi_{3}\rangle_{c,d} & =\frac{\alpha}{4}(\frac{\alpha^{2}-3r}{\sqrt{3}}(|3,0\rangle_{c,d}+e^{3i\varphi}|0,3\rangle_{c,d})+\\
 & (\alpha^{2}+r)(e^{i\varphi}|2,1\rangle_{c,d}+e^{2i\varphi}|1,2\rangle_{c,d}))\nonumber \end{align}

After the two mode state is mixed again at the second BS we can consider
the event of a 3 photon absorption in detector \textbf{f}:

\begin{align}
|\langle0,3|\psi_{3}\rangle_{e,f}|^{2} & =\frac{|\alpha|^{2}}{128}|(\frac{(\alpha^{2}-3r)}{\sqrt{3}}(1-e^{3i\varphi})+\\
 & \sqrt{3}(\alpha^{2}+r)(e^{2i\varphi}-e^{i\varphi})|^{2}\nonumber \end{align}

The boundary cases here are of special interest; for a purely classical
input ($r=0$)

\begin{equation}
|\langle0,3|\psi_{3}\rangle_{e,f}|_{classical}^{2}=\frac{|\alpha|^{6}}{12}\sin^{6}(\frac{\varphi}{2})\end{equation}

Indeed, 3-photon absorption of a classical interference pattern corresponds
to the 3rd power of linear absorption phase dependence, $\sin^{2}(\frac{\varphi}{2})$.
Equal amounts of quantum and classical light ($\alpha^{2}=-r$ ) would
give here the 3-photon NOON state, which would have the following
dependence on $\varphi$:

\begin{equation}
|\langle0,3|\psi_{3}\rangle_{e,f}|_{NOON}^{2}=\frac{|\alpha|^{6}}{6}\sin^{2}(\frac{3\varphi}{2})\end{equation}

\begin{figure}
\centering \includegraphics[clip,width=7cm]{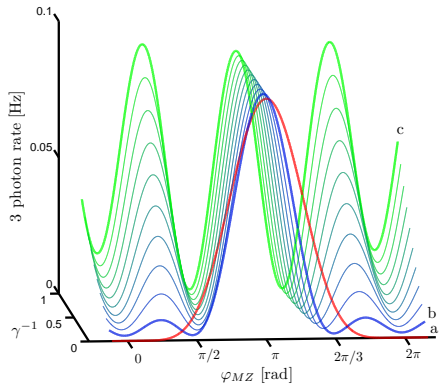} \caption{(color online). Simulation of 3-photon coincidence rate for $\gamma^{-1}=0,0.1,...,1$.
Plot \textbf{c} ($\gamma^{-1}=1$) corresponds to a 3-photon NOON
state displaying perfect super-resolving sinusoidal fringes. Remarkably,
it turns out that increasing $\gamma$, i.e. attenuating the non-classical
light, enables to selectively suppress certain peaks while leaving
others almost unchanged. When the SPDC is completely blocked (plot
\textbf{a}), the classical intensity phase dependence is obtained.
Conveniently, most of the quantum narrowing is achieved with minute
amounts of SPDC (plot \textbf{b}). }

\label{simulation}
\end{figure}

The width of the fringes in this case, matches that of classical light
with a wavelength that is three times shorter. The simulation in Fig.\ref{simulation}
illustrates the intermediate cases. It displays the dependence of
the 3-photon coincidence rate on the Mach Zehnder phase for different
values of $\gamma$. Most importantly, the losses within the system
and the efficiency of the detectors were taken into consideration.
In particular, this amounts to accounting for the effect of higher
coincidence orders on lower ones. The first plot (\textbf{a}), corresponding
to the purely classical state (no SPDC), displays the classical $\sin^{6}(\frac{\varphi}{2})$
dependence. As more quantum light is inserted into the system, the
central peak becomes narrower and the plot gradually turns into the
3-photon NOON state (plot \textbf{c}). Remarkably, most of the narrowing
effect of the central feature occurs at very low levels of SPDC. As
an example of this, the case of $\gamma^{-1}=0.1$ is highlighted
(plot \textbf{b}). In this example, 67\% of the maximum possible narrowing,
is achieved with only 10\% of the quantum light required for creating
perfect NOON states.

\begin{figure}
\centering \includegraphics[width=7cm]{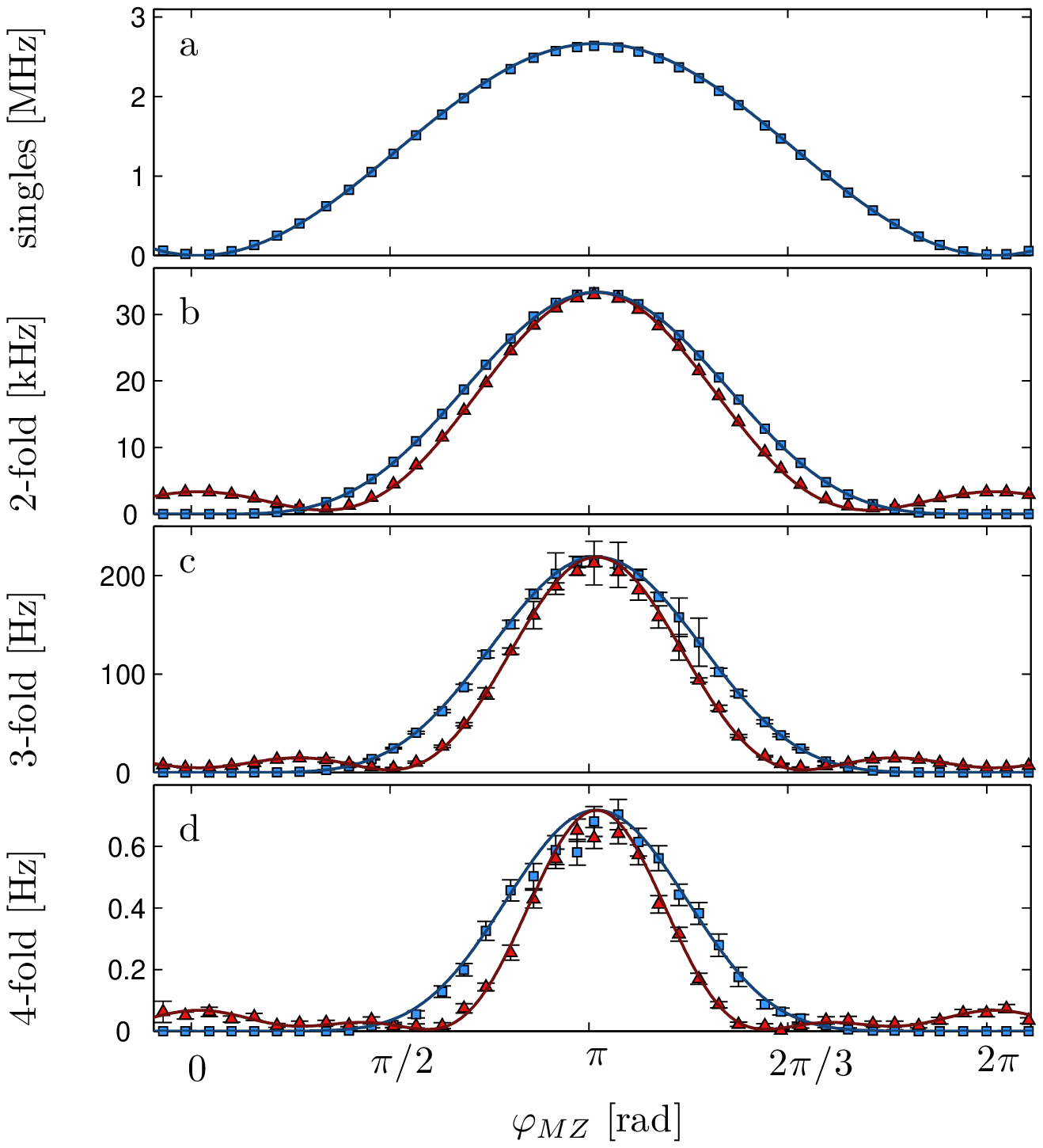} \caption{(color online). Experimental N-fold coincidence measurements for mixing
SPDC and CS with a pair ratio of 1:10 ($\gamma^{-1}=0.1$) (\textbf{red
triangles}) shown together with classical light only (\textbf{blue
squares).} Solid lines are obtained from a simulation with no free
parameters which takes into account the loss in the system and the
detection efficiencies. \textbf{a} Classical single counts demonstrate
classical resolution with a $\sin^{2}(\frac{\varphi}{2})$ intensity
dependence. Boxes \textbf{b-d} display number of $|N\rangle_{e}|0\rangle_{f}$
events, i.e. N simultaneous {}``clicks'' in detector \textbf{e},
with N=2-4 respectively. With only 10\% of the quantum light required
for a perfect NOON state, we create features with a width that scales
like N$^{-0.7}$. In the 3-photon case, this amounts to 67\% of the
maximum possible narrowing, which is obtained with perfect NOON states.}

\label{results}
\end{figure}

\emph{Experimental setup and results}. The experimental setup is similar
to the one described in \cite{afek} and is shown in Fig.\ref{setup}(B).
Briefly, the scheme begins with generation of SPDC and coherent light
with shared spatial and spectral modes in perpendicular linear polarizations
($\hat{H}$,$\hat{V}$). The relative phase between the beams is adjusted
with a liquid crystal (LC) retarder after which they are overlapped
on a polarizing beam splitter (PBS) cube. The phase shifter (PS) in
the Mach-Zehnder (MZ) interferometer is implemented with the use of
another LC oriented in $45^{0}$ to the linear polarizations. This
ensures phase stability in a co-linear, polarization based MZ. The
two modes of the light are subsequently coupled into a polarization
maintaining fiber which achieves the spatial mode matching. The output
modes ($\hat{H}$,$\hat{V}$) are separated by another PBS and detected
by a photon number-resolving detection apparatus which is composed
of an array of single-photon avalanche photodiodes. N photon events
in port $\hat{f}$ are measured as a function of the MZ phase ($\varphi$).
We preform this measurement for $N=1,2,3,4$, (Fig.\ref{results}).
For $N\geq2$, the quantum result (at $\gamma^{-1}=0.1$) is shown
together with the classical ($\gamma^{-1}=0$) N-photon absorption.
The additional narrowing due to the SPDC is clearly visible.

\begin{figure}
\centering \includegraphics[width=7cm]{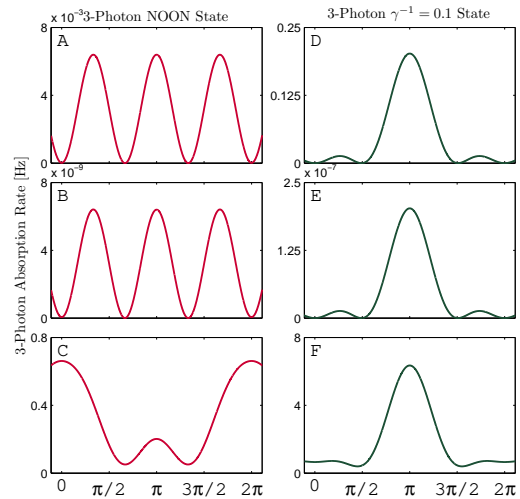} \caption{(color online). Simulation of 3-photon coincidence rates for NOON
state (left column) and for a $\gamma^{-1}=0.1$ state (right column).
The simulation takes into account losses in the system and imperfections
in the quantum states due to high input photon flux. We define the
setup transmission $\eta$ as the ratio of the coincidences to singles
in the SPDC, and the quantum flux $N_{p}$ as the rate of pairs in
the SPDC. The figures in the top row ($A$ and $D$) show the ideal
case in which the quantum states are flawless due to low photon flux
($N_{p}=10^{-6}Hz$) and perfect transmission ($\eta=1$). The next
row ($B$ and $E$) illustrates the more practical case for lithography
whereby the limited absorption of an N-photon resist would greatly
reduce the efficiency ($\eta=0.01$). In this case, leaving the photon
flux unchanged ($N_{p}=10^{-6}Hz$) leads to virtually the same functional
dependence at the cost of very low 3-photon detection rate. In order
to reach a 3-fold coincidence rate which would be usable in practical
lithography the flux must be substantially increased ($N_{p}=0.1Hz$)
as shown in the last row ($C$ and $F$). Since the approximation
under which the NOON state was created is no longer valid at such
fluxes it is completely deformed ($C$) while the $\gamma^{-1}=0.1$
state remains all but unharmed ($F$). }

\label{results2}
\end{figure}

The simulation in Fig. \ref{results2} compares the performances of
a $\gamma^{-1}=0.1$ state with those of a NOON state under various
experimental conditions. It shows the phase dependence of the 3-fold
coincidence under ideal experimental conditions for both the NOON
state ($A$) and the $\gamma^{-1}=0.1$ state ($D$). The next row
($B$ and $E$) shows that under conditions that greatly reduce efficiency,
the detection rate is drastically attenuated but its phase dependence
is unchanged provided that the input flux remains the same. Clearly,
in order to render the 3-photon flux practical for multi-photon lithography,
the input flux must be increased. The bottom row shows the two states
under the same efficiency conditions but with higher input fluxes.
The NOON state is distorted under these conditions as the approximations
under which it was created are no longer valid ($C$) , \cite{Hofmann}.
At the same time, the $\gamma^{-1}=0.1$ state remains essentially
unchanged ($F$).

\emph{Conclusion}. We conclude by saying that a major impediment of
quantum lithography, due to which it was hitherto deemed impractical,
comes from the limited intensity of quantum light sources. In the
scheme demonstrated here only $10\%$ of the photon pairs are of quantum
origin (originally entangled) thus allowing high fluxes of photons.

Moreover, it is safe to presume that any multi-photon photo-resist
that would allow in the future practical quantum lithography would
absorb with an efficiency that is far from perfect, \cite{Boyd,Fourkas},
implying that detectable absorption rates would require an increased
generating flux. While current schemes for creating NOON states are
limited to low input fluxes, high flux loss-resistant states possess
classical behavior in the sense that higher input fluxes effect mostly
the intensity of the signal, not so much distorting its quality.

Financial support of this research by the ERC grant QUAMI, the Minerva
Foundation, and the Crown Photonics Center is gratefully acknowledged.

\end{document}